\documentclass[]{article}

\usepackage{latexsym}      
\usepackage{graphicx}      
\usepackage{amsmath}
\usepackage{subfigure}
\usepackage[]{hyperref}

\title{Analysis of the damped quantum search and its application to the one-dimensional Ising system}
\author{Neris Ilano, Cristine Villagonzalo and Ronald Banzon\\
\vspace{0.015in}
\textit{\footnotesize National Institute of Physics, University of the Philippines, }\\
\textit{\footnotesize Diliman, Quezon City, 1101 Philippines}}

\begin{document}

\maketitle

\begin{abstract}
An analysis on the damped quantum search by exploring the rate at which the target state is obtained. The results were compared with that of the classical search since the standard Grover's algorithm does not give a convergent result if the number of target state is unknown. For a large number of target states, the classical and the damped quantum search give a similar result. However, for intermediate values of the target size the damped quantum search gives a higher probability of success than the classical search. Furthermore, we also made an analysis on the average number of iterations needed to obtain at least one of the target states. As the number of target states is reduced, the damped quantum search gives a better result than the classical search. The results coincide if the size of target state is comparable to the size of the sample.
\end{abstract}

\section{Introduction}\label{intro}
Harnessing the power of quantum mechanics in computation promises a new era in computer science. Two well known quantum algorithms that shows superiority over their classical counterparts are due to Shor and Grover. Shor demonstrated that the problem of finding the prime factors of an integer can be solved efficiently on a quantum computer \cite{Shor}. Grover showed that the problem of conducting a search through some unsorted database can be sped up on a quantum computer \cite{Grover0}. While Grover's algorithm did not provide an astonishing speed up as Shor's algorithm, the widespread applicability of the search-based problem has excited significant interest in Grover's algorithm \cite{Nielsen}.

We can describe the search problem as follows: Consider a database with $N$ unsorted objects, among which $M$ of them are the desired objects, how many times on the average do we need to search in order to find the first desired object? Using classical computation it will take $N/M$ search on the average before the first desired object is obtained. A detailed derivation of this result is in \cite{Brylinsky}. If we use the quantum search algorithm it will take $\sqrt{N/M}$ on the average before we obtained the first desired object \cite{Grover0}. Experimental implementations of Grover's algorithm on a small scale quantum computer are reported in \cite{Chuang, Kim} and they were able to show that the search requires fewer steps than on a classical computer.

The introduction of the Grover's search algorithm spark the interest of some researchers to address its shortcomings \cite{Grover0, Grover1, Grover2, Tulsi}. It has been established already that the Grover's algorithm provides a quadratic speed up over its classical counterpart and was shown to be the best oracle based search algorithm \cite{Boyer}. However, the Grover's algorithm is not without limitations. One of these is knowing the number of target states before the procedure is performed. Without such knowledge, the process simply oscillates. One way to approach this dilemma is to introduce a damping parameter to suppress the mentioned oscillation. This was done by Mizel in \cite{Mizel} where he attached an external spin with the corresponding damping parameter. He demonstrated that there exists a critical damping which separates the classical result from quadratic speed up of Grover's result.

The quantum search algorithm \cite{Grover0} starts from an initial state vector to a target state with a number of iterations. This leads to a quadratic speed up over its corresponding classical algorithm in searching an unsorted database. To have an optimum result, one needs to know the number of target states. 
Boyer \textit{et al.} provided a tight analysis of Grover's quantum search algorithm by giving a simple closed-form formula for the probability of success after any given number of iterations \cite{Boyer}:
\begin{eqnarray}\label{closed-form}
	k_j&=\frac{1}{\sqrt{M}}\sin((2j+1)\theta)  \\
	l_j&=\frac{1}{\sqrt{N-M}}\cos((2j+1)\theta)
\end{eqnarray}
where $k$ and $l$ are the probability amplitudes of the target and nontarget states respectively. Here $M$ is the number of the target states, $N$ is the total number of items in our database, $j$ is the number of iterations, and the angle $\theta$ is defined so that $\sin^{2}\theta=M/N$. This allows us to determine the number of iterations necessary to achieve with almost certainty of finding the target state, as well as the upper bound on the probability of failure. The closed-form formula for $k_j$ and $l_j$ could leisurely be derived by mathematical induction.

Mizel proposed a quatum search algorithm that utilizes dissipation in order to make the search robust even without the knowledge of the number of target states \cite{Mizel}. However, providing a closed-form formula for the case of damped quantum search is not trivial.  

In this work, we apply the damped quantum search in an unsorted database of 8- and 12-spin Ising systems and provide an analysis of the damped quantum search. 

\section{Damped Quantum Search in an Ising Spin System}\label{DQS}
The recent quantum search algorithm introduced by Mizel \cite{Mizel} is appropriate in searching an unsorted database without the prior knowledge of the number of target items $M$. We expound the significance of Mizel's algorithm by applying it to a one-dimensional Ising spin system database. We search for an eigenstate with 8 and 12 spins using the damped quantum search algorithm.  

An $n$-spin Ising system models a ferromagnet with $n$ magnetic dipole moments. Its Hamiltonian is given by $H=-\epsilon\sum_{i=1}^{n-1}s_{i}s_{i+1}$, where $\epsilon$ is the interaction energy. It can be shown that the Hamiltonian is a $2^{n}\times 2^{n}$ diagonal matrix. Here, $s$ is the $\sigma_{z}$-Pauli spin  matrix defined in a two-dimensional Hilbert space spanned by $|\uparrow\rangle$ and $|\downarrow\rangle$, representing the spin-up and spin-down states. 
Thus, we can construct an oracle that behaves as follows:
\begin{equation}\label{U_f}
U_{f}=
\begin{cases}
-1 &if H_{i,i}=\lambda  \\
+1 &if H_{i,i}\neq\lambda\\
\end{cases}
\end{equation}
where $\lambda$ is the desired energy eigenvalue.

We seek on 8- and 12-spin Ising systems for at least one of their eigenstates of a certain eigenvalue by applying the damped quantum search. For comparison, we also show the classical result and the quantum search result assuming ignorance of the degeneracy $M$. There are $N=2^{8}=256$ distinct configurations that we take as items in the database of the $8$ spins system, and $N=4096$ for the $12$ spins. Our task is to locate at least one of the eigenstates that corresponds to a certain eigenvalue, $\lambda$, for $n=8$ and $n=12$.

\subsection{Numerical Computation}\label{sec:numcomp}
To implement the damped quantum search algorithm in an $n$-spin Ising system, we find the degeneracy $M$ of the eigenvalue corresponding to $\lambda$ for validation purposes. It would be convenient for computational intent to write the Hamiltonian as follows:
\begin{equation}
{H}=-\epsilon\left[{S}\otimes{I}^{\otimes (n-2)}+\sum_{i=1}^{n-3}{I}^{\otimes (i)}\otimes{S}\otimes{I}^{\otimes(n-i-2)}
+{I}^{\otimes (n-2)}\otimes{S}\right],
\end{equation}
where $S=s\otimes s$ and $I=I^{\otimes 1}$ is a $2\times 2$ identity matrix,
\begin{equation}
S=\left(\begin{array}{cc}
  1 & 0 \\
  0 & -1
  \end{array}\right),
\qquad
I=\left(\begin{array}{cc}
  1 & 0 \\
  0 & 1
\end{array}\right).
\end{equation}
Since $s$ and $I$ are all diagonal matrices, then $H$ is also diagonal and finding the eigenvalues is straightforward. 

Let us now implement the quantum search algorithm. We construct the initial state $|\psi\rangle$ and the Grover rotation operator ${G}$ using our knowledge of $N=2^{n}$ and the degeneracy of the desired $\lambda$, $M$,
\begin{equation}\label{eq:g}
G=\frac{N-2M}{N}
	\left(\begin{array}{cc}
  1 & 0 \\
  0 & 1
  \end{array}\right)
	+2\sqrt{\frac{NM-M^2}{N^2}}
	\left(\begin{array}{cc}
	1 & 0 \\
	0 & -1
\end{array}\right)
\end{equation}  
The probability that the target state has been reached were recorded in each iteration. 

On the other hand, for the damped quantum search we assume ignorance of the $M$. We start with the construction of the square matrix 
\begin{equation}\label{eq:mat}
\left(\begin{array}{c}
	\mathbf{Tr}(\rho'X) \\
	\mathbf{Tr}(\rho'Z) \\
	\mathbf{Tr}(\rho')  \\
	\end{array}\right)
	=
	\left(\begin{array}{ccc}
		\cos2\theta\cos\phi &\sin2\theta\frac{1+\cos^{2}\phi}{2} &\sin2\theta\frac{1-\cos^{2}\phi}{2} \\
	 -\sin2\theta\cos\phi &\cos2\theta\frac{1+\cos^{2}\phi}{2} &\cos2\theta\frac{1-\cos^{2}\phi}{2} \\
	  0 									&\frac{1-\cos^{2}\phi}{2} 					 &\frac{1+\cos^{2}\phi}{2} 						\\
	\end{array}\right)
	\left(\begin{array}{c}
	\mathbf{Tr}(\rho X) \\
	\mathbf{Tr}(\rho Z) \\
	\mathbf{Tr}(\rho)   \\	
	\end{array}\right)
\end{equation}
where $\rho=|\psi\rangle\langle\psi|$, 
$X=\left(\begin{array}{cc}
	0 & 1\\
	1 & 0
	\end{array}\right)$ 
and 
$Z=\left(\begin{array}{cc}
	1 & 0\\
	0 & -1
	\end{array}\right)$. The damping parameter $\phi$ is defined as $\cos\phi=(1-\sin\theta)/(1+\sin\theta)$. Equation \ref{eq:mat} will be applied on the column vector
\begin{equation}
	\left(\begin{array}{c}
		\mathbf{Tr}(\rho X) \\
		\mathbf{Tr}(\rho Z) \\
		\mathbf{Tr}(\rho)   \\	
	\end{array}\right)
	=
 			\left(\begin{array}{c}
 			\sin\theta\\
 			\cos\theta\\
 			1
 			\end{array}\right)
\end{equation} 
The value of the third entry of the resulting column vector is then subtracted from one and the result is plotted as a function of iteration.

\subsection{Probability of Success}
\begin{figure}[!htp]\label{fig:grover121}
\centering
\mbox{\subfigure[$M=22$]{\includegraphics[height=1.9in, width=0.5\columnwidth]
	 					{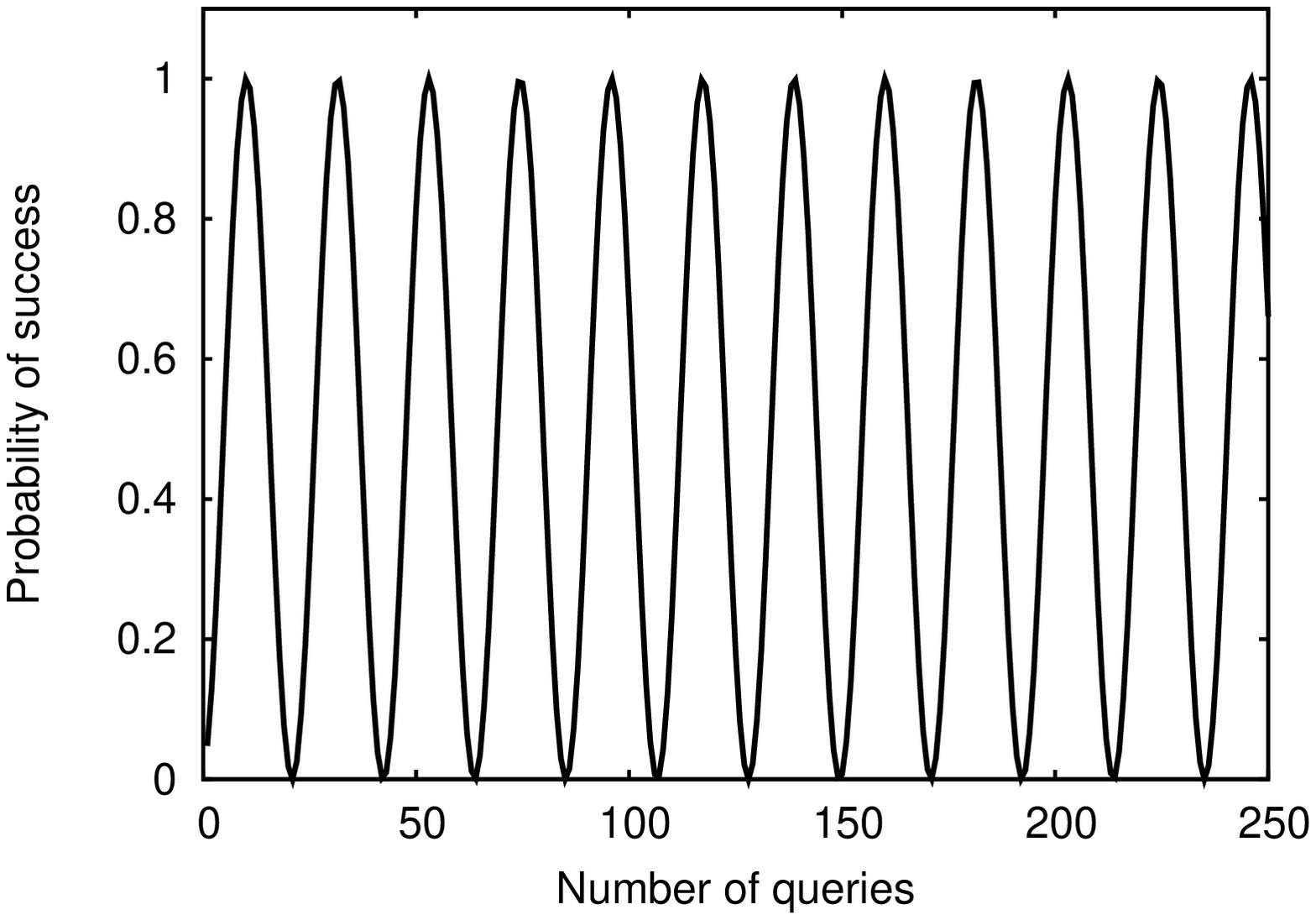}}\quad
	  \subfigure[$M=22$]{\includegraphics[height=1.9in, width=0.5\columnwidth]
	 					{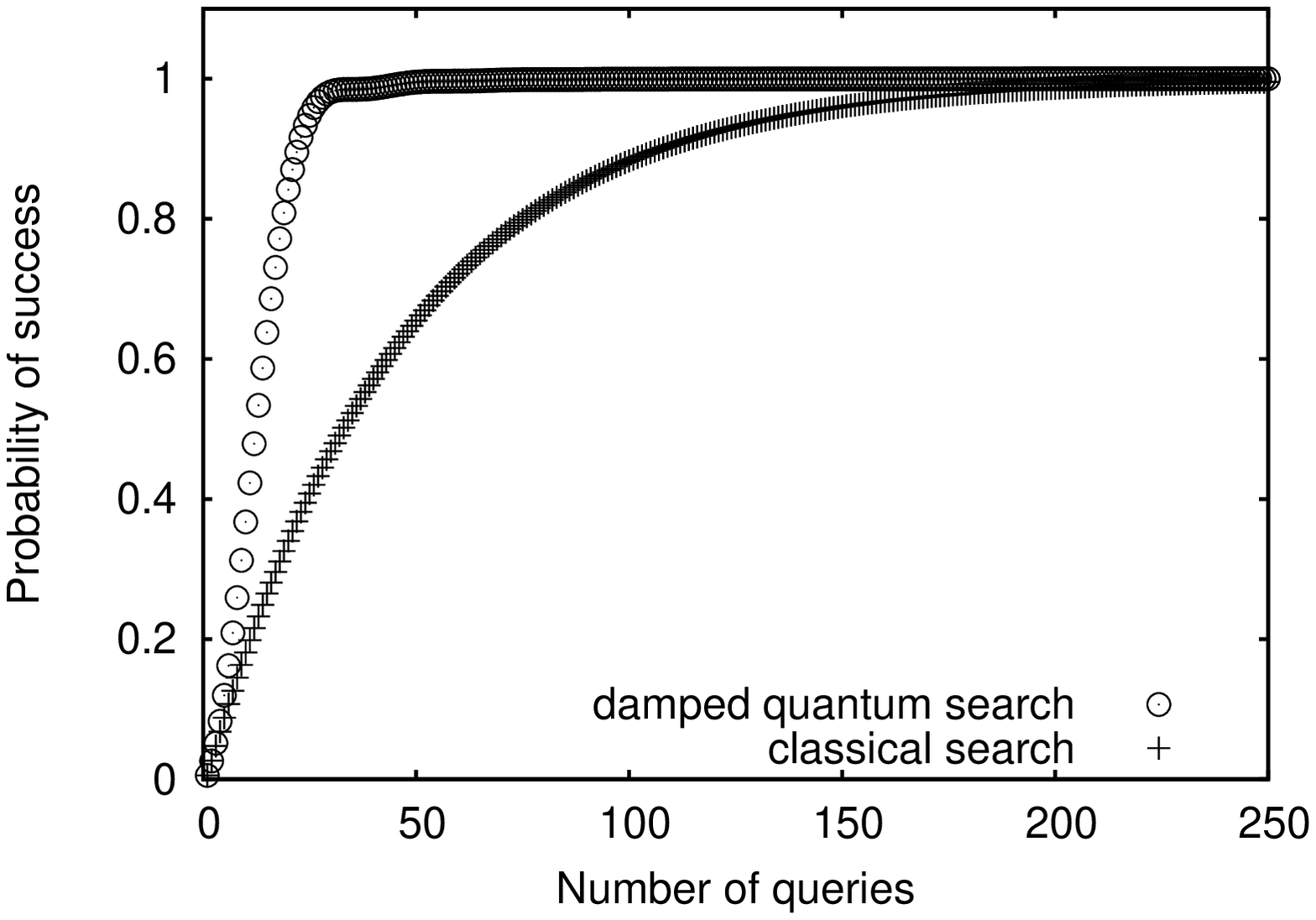}}} 
\mbox{\subfigure[$M=110$]{\includegraphics[height=1.9in, width=0.5\columnwidth]
	 					 {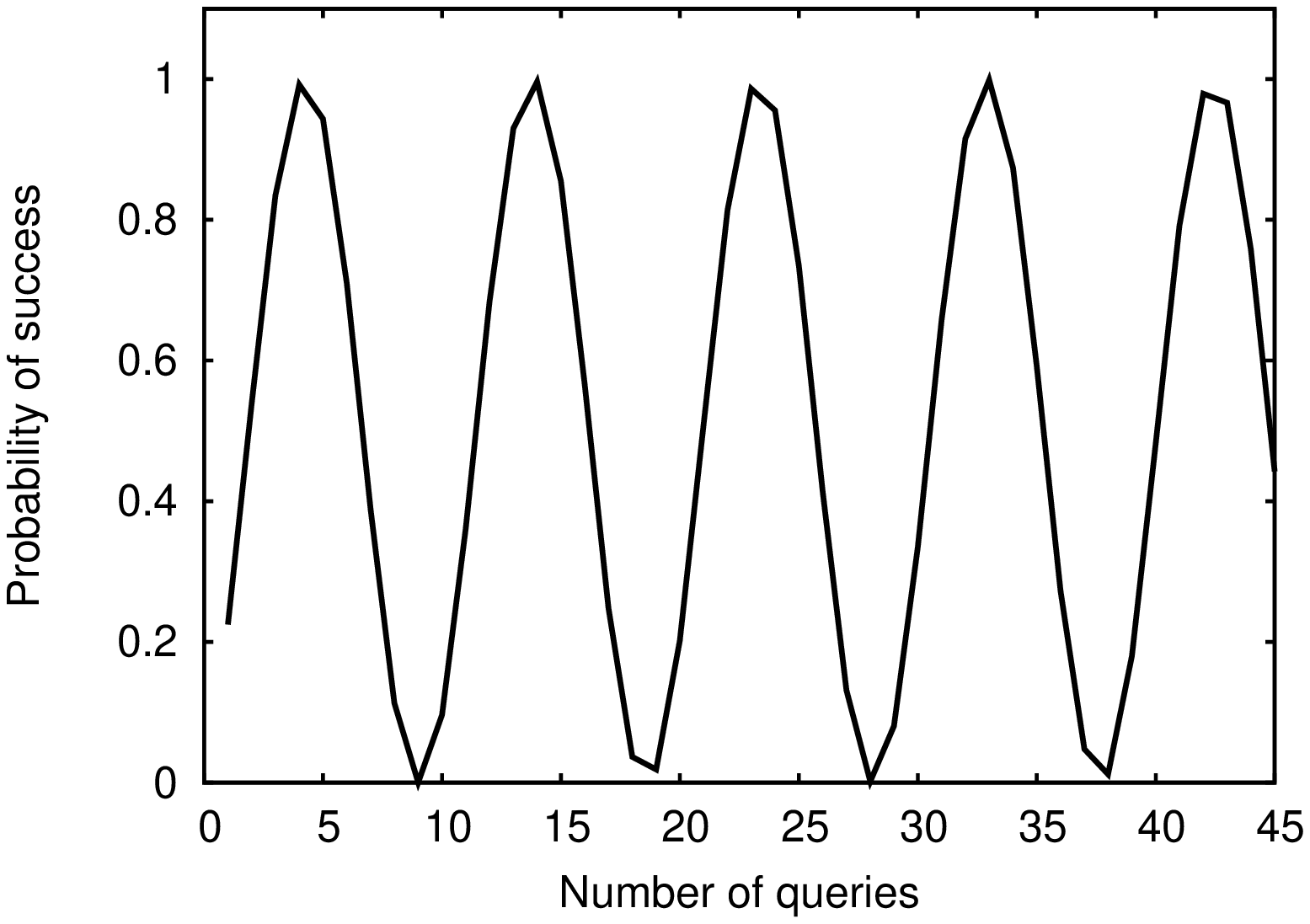}}\quad
	  \subfigure[$M=110$]{\includegraphics[height=1.9in, width=0.5\columnwidth]
	 					 {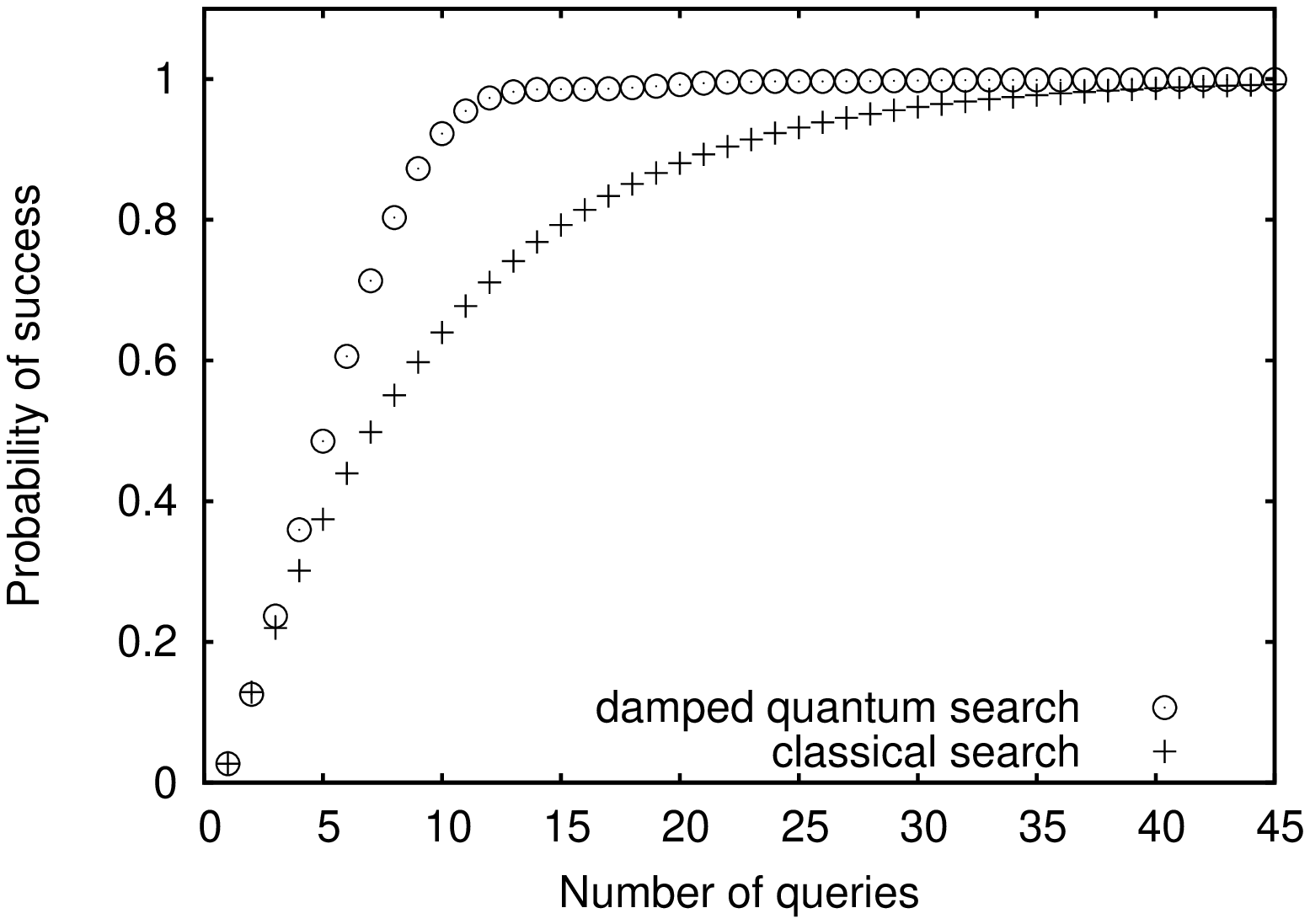}}}
\caption{(a) and (c): The oscillations of the probability of finding the target state using quantum search, (b) and (d): the increase in the probability that the final system will collapse to $|\beta\rangle\langle\beta|\otimes|\uparrow\rangle\langle\uparrow|$ using damped quantum search and classical search for $N=4096$ at a given $M$.}
\label{fig:prob2}
\end{figure}

Figure \ref{fig:prob2} shows the behavior of the probability of finding the target state for a given $M$ in an 8-spin Ising system using the Grover's quantum search, classical search and damped quantum search as the number of queries increases.

The oscillations of the probability is evident in Grover's quantum search. If we do not know the degeneracy of the desired eigenvalue, then the search is not optimized. Furthermore, in the damped quantum search, the probability that the target state will be found and the external spin flipped steadily approach unity as we increase the number of queries. This is similar to the classical limiting case. 
The advantage of the damped quantum search is that it preserves the 

\noindent signature $O(\sqrt{M/N})$ number of queries \cite{Mizel}. Consider for example Figure~\ref{fig:prob2} (a), for $n=12$ with $M=22$, there are $10$ iterations needed to make Grover's search robust. This gives a probability of $99.9\%$. The damped quantum search locates one of the target states after $32$ queries with a probability of $99.5\%$ as shown in Figure~\ref{fig:prob2} (b). Ignorance of the degeneracy requires additional queries but are minimal compared to the classical search. The latter takes $247$ queries with the same probability.

\section{On the Average Number of Iterations}
To generalize our results, we employ the concept of the average number of iterations. We start with $j$ iterations. After the $j^{th}$ iteration, we check if the target state has been found. If the result is positive, we stop the search. Otherwise, in case of failure we restart the iterations. After the first $j$ iterations has failed, the average number of iterations, $E(j)$, is given by
\begin{equation}\label{eq:Ej}
	E(j)=\frac{j}{P(j)}
\end{equation}
where $P(j)$ is the probability that the target has been found after the $j^{th}$ iteration. In line with this, we want to obtain a certain value of $j$ that will give the minimum average number of iterations, $E_{csmin}$ for the classical search and  $E_{dqsmin}$ for the damped quantum search, before success. If a closed form of $P(j)$ is available, then we calculate $E'(j)$ and equate it to zero. This is not the case for the damped quantum search since the closed form of $P(j)$ cannot be obtained easily and the form of $P(j)$ varies for each iteration. Therefore, we can plot $E(j)$ with respect to $j$ using \ref{eq:Ej} to obtain the minimum graphically . A summary of all minimum average number of iterations before success for 8- and 12-spin Ising system are summarized in Tables \ref{tab:minave8} and \ref{tab:minave12}, respectively. The overhead in the number of iterations of the classical search relative to the damped quantum search is obtained from their ratio.

\begin{table}
\centering
\caption{\label{tab:minave8} Minimum expected number of iterations before success for classical search and damped quantum search. We take $N=256$ for 8 spins and assume ignorance of the degeneracy $M$ for a given energy $\lambda$.}
\begin{tabular}{@{}l|l r @{.} l r @{.} l r @{.} l @{}}
	 $\lambda$ &$M$ &\multicolumn{2}{c}{$E_{csmin}$}   &\multicolumn{2}{c}{$E_{dqsmin}$}   &\multicolumn{2}{c}{$E_{csmin}/E_{dqsmin}$} \\
	 \hline\hline
	$\pm 7\epsilon$	  &2     &39&1796     &19&1233      &2&04879\\	
	$\pm 5\epsilon$   &14    &7&4034      &6&5365       &1&13260\\
	$\pm 3\epsilon$   &42    &3&2121      &3&2833       &0&97831\\
	$\pm 1\epsilon$   &70    &2&3507      &2&3986       &0&98003\\
\end{tabular}
\end{table}

\begin{figure}[!htp]
\centering
\mbox{\subfigure[$M=14$]{\includegraphics[height=1.9in, width=0.5\columnwidth]
	                    {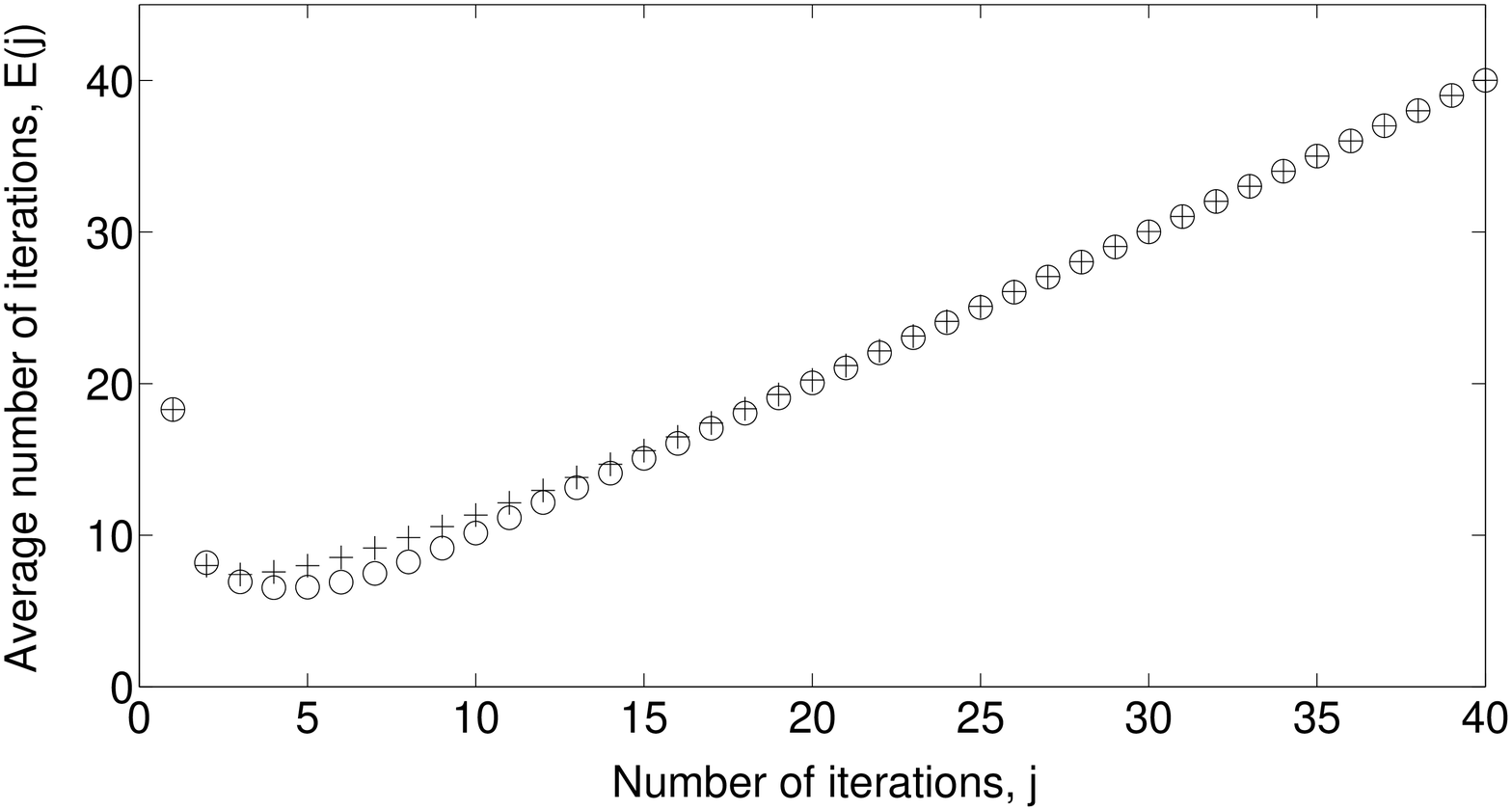}}\quad
	  \subfigure[$M=42$]{\includegraphics[height=1.9in, width=0.5\columnwidth]
	                    {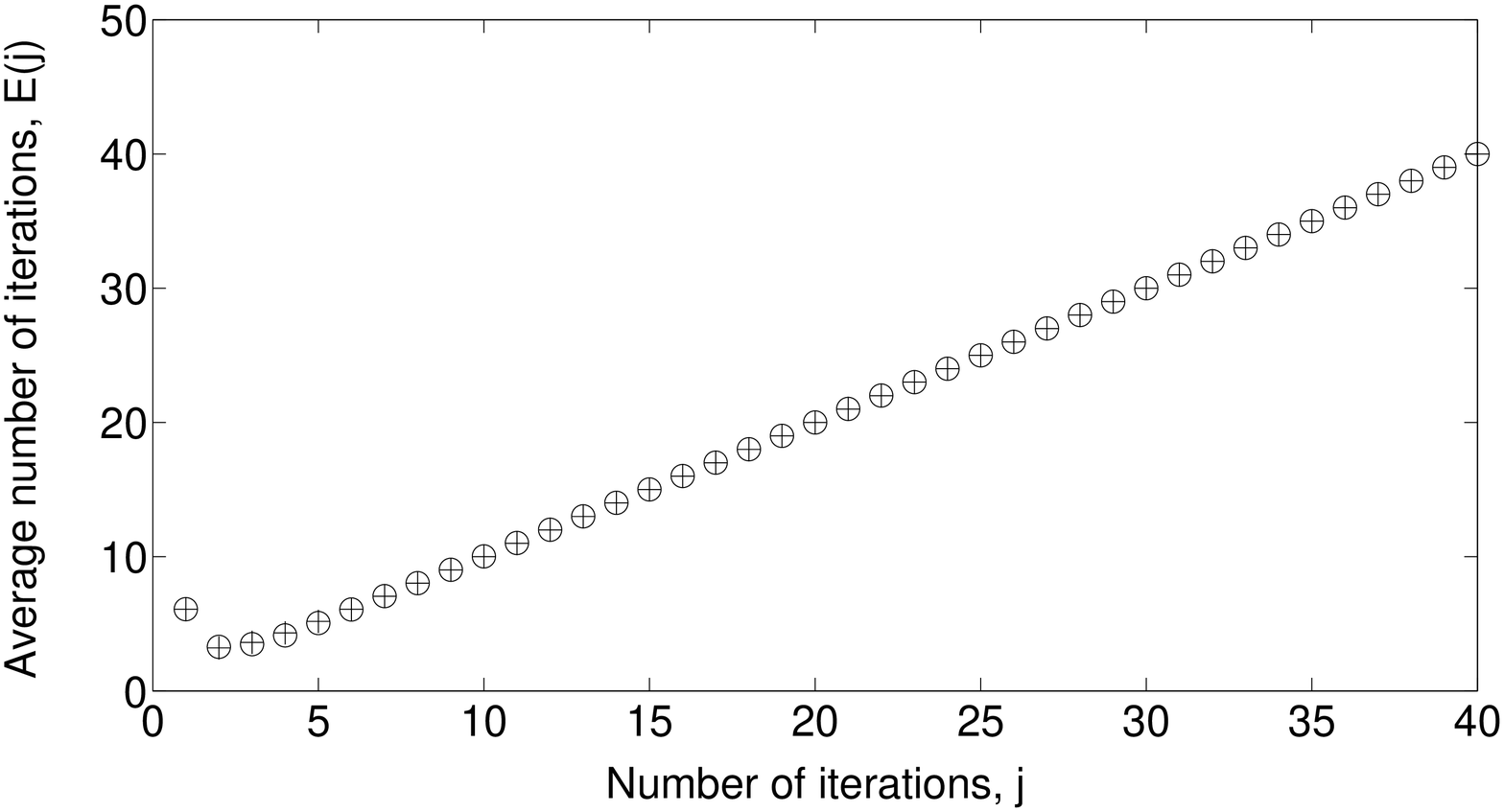}}}
\caption{Expected number of iterations before success of classical search ($+$) and damped quantum search ($\circ$) as a function of the number of iterations for $N=256$ at a given $M$.}
\label{fig:ave8}
\end{figure}
\begin{figure}[!htp]
\centering
\mbox{\subfigure[$M=110$]{\includegraphics[height=1.9in, width=0.5\columnwidth]
						 {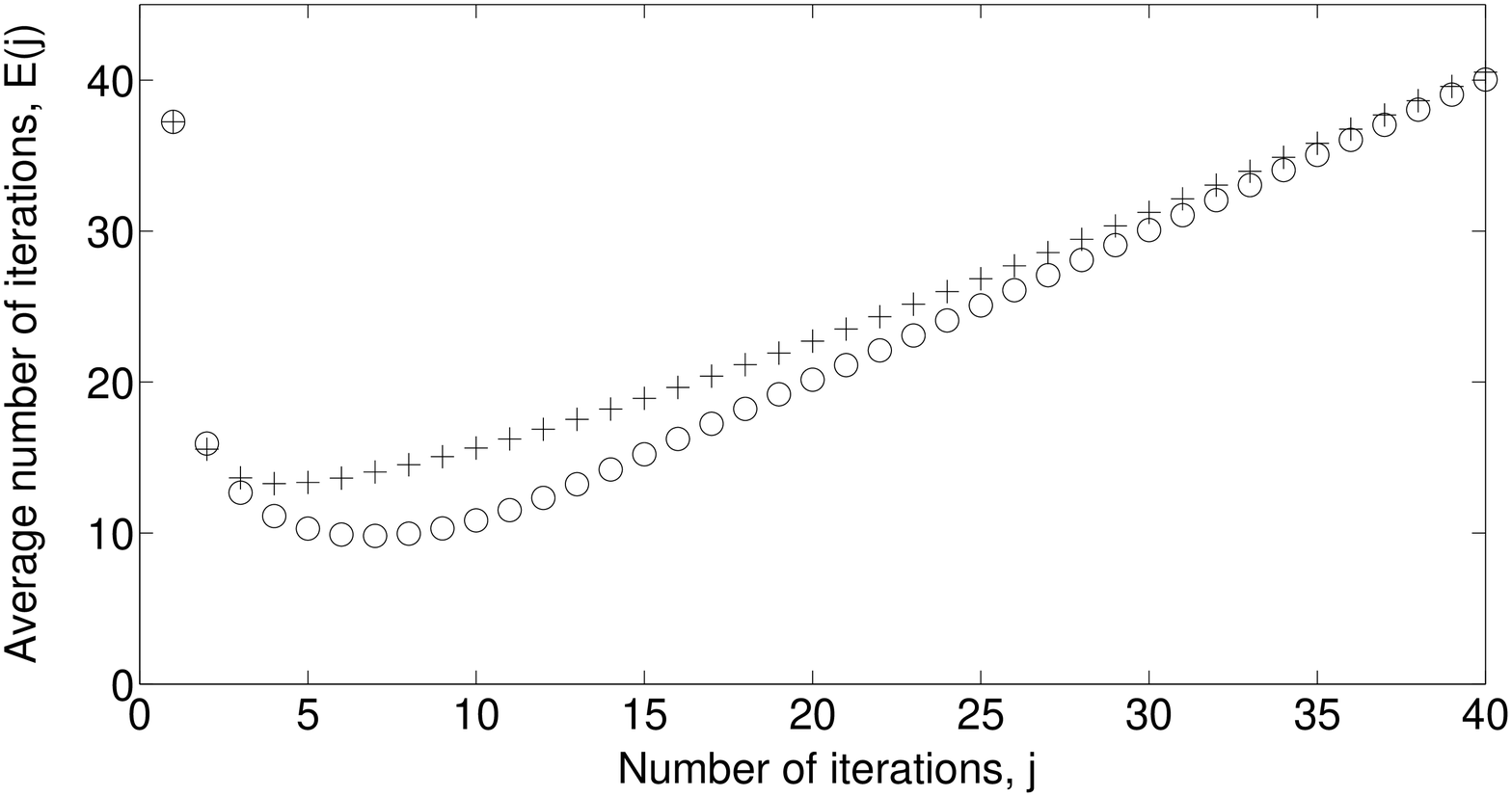}}\quad
	  \subfigure[$M=660$]{\includegraphics[height=1.9in, width=0.5\columnwidth]
						 {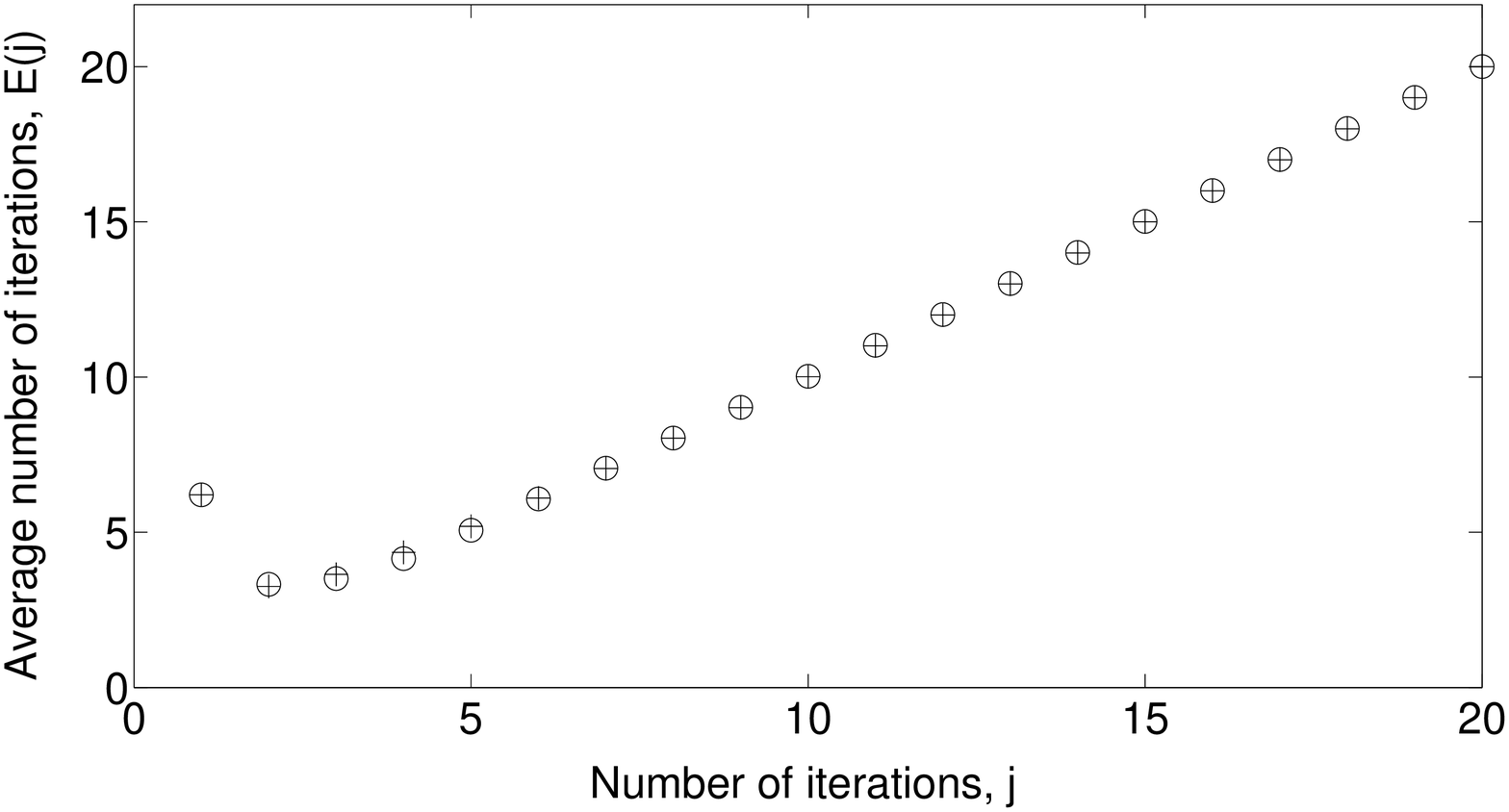}}}
\caption{Expected number of iterations before success of classical search ($+$) and damped quantum search ($\circ$) as a function of the number of iterations for $N=4096$ at a given $M$.}
\label{fig:ave12}
\end{figure}

\begin{table}
\centering
\caption{\label{tab:minave12} Minimum expected number of iterations before success for classical search and damped quantum search. We take $N=4096$ for 12 spins and assume ignorance of the degeneracy $M$ for a given energy $\lambda$.}
\begin{tabular}{@{}l|l r @{.} l r @{.} l r @{.} l@{}}
	 $\lambda$  &$M$   &\multicolumn{2}{c}{$E_{csmin}$}   &\multicolumn{2}{c}{$E_{dqsmin}$}   &\multicolumn{2}{c}{$E_{csmin}/E_{dqsmin}$}\\
	 \hline\hline
	$\pm 11\epsilon$	&2        &539&9602	   &79&2205    &6&8159\\	
	$\pm 9\epsilon$     &22       &55&1413     &23&2660    &2&3700\\
	$\pm 7\epsilon$	    &110      &13&2779	   &9&8111     &1&3533\\	
	$\pm 5\epsilon$     &330      &5&5076      &5&1825     &1&0627\\
	$\pm 3\epsilon$     &660      &3&2537      &3&3259     &0&9784\\
	$\pm 1\epsilon$     &924      &2&6085      &2&6638     &0&9792\\
\end{tabular}
\end{table}
Using the energy eigenvalues as the search marker of the target states makes the search process dependent on the corresponding degeneracies. The behavior of the average number of iterations is the same for both the classical and the quantum search. However, the difference of the classical and the quantum search is evident when the target state is one of the highly excited state where the degeneracy is very small. The ratio of the average number of iterations approaches unity as the energy eigenvalues, which serve as the marker, approaches the ground state energy. This shows that the quantum search process approaches the classical result if there are several states with the same marker.

\section{Conclusion}
We have shown the advantage of damping the quantum search. Application of the Grover's quantum search in an Ising system reveals its oscillatory nature that leads to a frail search. This dilemma becomes more evident if we are unaware of the degeneracy $M$ in advance. The damped quantum search provides a fixed point of convergence, that is, increasing the number of queries will get us closer to the target state. Furthermore, for large number of target states, the damped quantum search transitioned to classical search. We consider this as an advantage in quantum searching because the number of iterations needed by the search algorithm increases with $M$ for $M\geq N/2$, thus we prefer using classical search in this case. On the other hand, the overhead on the minimum number of iterations before success of the damped quantum search over the classical search is apparent for small number of target states. 

\section*{Acknowledgement}
N.I. acknowledges support from the Department of Science and Technolohy SEI-ASTHRDP.


\begin{thebibliography}{100}
\bibitem{Shor} 
Shor P 1994 
{\it Proc. 35th Annual Symp. on Fundamentals of Computer Science} 
(Los Alamitos, CA: IEEE Press) pp~124--34
\bibitem{Grover0}
Grover L K 1997 
{\it Phys. Rev. Lett.} {\bf 78} 325–-28
\bibitem{Nielsen} 
Nielsen M A and Chuang I L 2000 
{\it Quantum Computation and Quantum Information} 
(Cambridge: Cambridge University Press) 
\bibitem{Brylinsky} 
Brylinsky R and Chen G 2002 
{\it Mathematics of Quantum Computation} 
(Florida: Chapman and Hall/CRC)
\bibitem{Chuang} 
Chuang I L, Gershenfeld N and Kubinec M 1998 
{\it Phys. Rev. Lett.} {\bf 80} 3408--11
\bibitem{Kim} 
Kim J, Lee J and Lee S 2002 
{\it Phys. Rev. A} {\bf 65} 054301
\bibitem{Grover1} 
Grover L K 2005 
{\it Phys. Rev. Lett.} {\bf 95} 150501
\bibitem{Grover2} 
Grover L K, Patel A, Tulsi T 2006 
Quantum algorithms with fixed points: The case of database search {\it e-print} quant-ph/0603132
\bibitem{Tulsi} 
Tulsi T, Grover L K, Patel A 2006 
{\it Quant. Inform. Comput.} {\bf 6} 483--94
\bibitem{Boyer} 
Boyer M, Brassard G, H$\o$yer P and Tapp A 1998 
{\it Fortsch. Phys. - Prog. Phys.} {\bf 46} 493-505
\bibitem{Mizel} 
Mizel A 2009 
{\it Phys. Rev. Lett.} {\bf 102} 150501
\end{thebibliography}
\end{document}